\documentclass[10pt,onecolumn]{article}
\usepackage{graphicx}
\usepackage{amsmath}
\usepackage{geometry}
\usepackage{authblk}
\usepackage{multicol}
\usepackage{titlesec}
\usepackage{comment}
\usepackage{hyperref}
\usepackage{natbib}

\titleformat{\section}{\bfseries\normalsize}{\thesection.}{1em}{}
\titleformat{\subsection}{\itshape\normalsize}{\thesubsection.}{1em}{}

\title{\vspace{-1.5cm}\rule{\linewidth}{1pt}\\[0.5em]
\textbf{\LARGE Image-Based Biospeckle Contrast Analysis for Rapid Antimicrobial Susceptibility Testing}\\[0.25em]
\rule{\linewidth}{1pt}
}

\author[1]{M.A. Gameiro}
\author[1]{R.F. Pinto}
\author[1]{N.V. Lopes}
\affil[1]{\small Center for Innovative Care and Health Technology, Polytechnic University of Leiria, Leiria, Portugal}

\date{}

\begin{document}

\maketitle

\section*{Abstract}
\noindent
\textit{Purpose:} Antimicrobial resistance is a major global health concern, affecting hospital admissions and treatment success. This study aims to introduce an experimental setup for monitoring bacterial activity over time using image-based contrast as a key biomarker.\\
\textit{Methods:} The proposed method captures changes in bacterial activity by analyzing variations in image contrast. The approach is evaluated for its ability to detect antimicrobial effects over shorter time intervals compared to conventional clinical techniques.\\
\textit{Results:} The findings reveal a progressive decrease in contrast over time, suggesting its potential as an indicator of antimicrobial activity. The results highlight the method's capability for early detection of bacterial susceptibility.\\
\textit{Conclusion:} The study demonstrates that image-based contrast analysis can serve as a rapid and reliable tool for antimicrobial susceptibility testing, offering advantages over traditional methods in clinical practice.

\textbf{Keywords:} Biospeckle, antimicrobial susceptibility, bacterial activity, optical imaging, rapid testing

\section{Introduction}

Antimicrobial resistance (AMR) poses a serious threat to global health, recognized by the World Health Organization (WHO) as one of the top ten global health threats \cite{walsh2023antimicrobial}. In Intensive Care Units (ICUs), nosocomial infections represent 20-30\% of admissions, significantly increasing mortality rates and healthcare costs \cite{brusselaers2011rising, gamer1988cdc, bereket2012update}. The widespread use of broad-spectrum antimicrobials in more than 60\% of ICU patients fosters the emergence of multidrug-resistant (MDR) pathogens such as Methicillin-Resistant Staphylococcus Aureus (MRSA) \cite{gajic2022antimicrobial} and Pseudomonas/Acinetobacter \cite{oberhettinger2020evaluation}.

To combat AMR, the Center for Disease Control (CDC) outlines crucial strategies including enhanced diagnostic and treatment approaches and the prudent use of antimicrobials \cite{dubourg2018rapid, tansarli2020diagnostic}. Traditional Antimicrobial Susceptibility Testing (AST) methods, crucial for guiding effective antimicrobial therapy, are often time-consuming, delaying critical treatment decisions in acute cases like sepsis.

In response, emerging technologies such as optical microscopy and laser speckle analysis offer promising alternatives for rapid AST. This article introduces a novel automated system utilizing laser speckle contrast analysis to monitor microbial activity. By measuring the contrast changes in speckle patterns caused by bacterial growth and movement, this system provides a direct, quantitative measure of bacterial activity over time.

The ability to track these changes reliably translates into faster AST, as the system can potentially detect early signs of growth inhibition or proliferation in response to antimicrobial agents. This capability could dramatically shorten the time required to determine the susceptibility patterns of pathogens, allowing for more timely and targeted antimicrobial interventions \cite{schmitz2018genmark}.

By establishing contrast as a reliable measure of microbial activity, the research presented herein not only enhances our understanding of microbial dynamics but also paves the way for innovations in AST methodologies. This advancement is particularly crucial for improving outcomes in settings where rapid response to infections is essential \cite{charretier2016mass}.

This study introduces a novel approach to antimicrobial susceptibility testing using laser speckle imaging, addressing key limitations of traditional methods by providing a rapid and dynamic assessment of bacterial activity. Unlike established techniques such as disk diffusion, MIC tests, or MALDI-TOF, which are time-intensive or require specialized setups, this method offers real-time, quantitative insights through contrast changes in speckle patterns. The experimental results demonstrate significantly faster detection of bacterial growth inhibition, highlighting the system's potential to revolutionize diagnostics and improve outcomes in combating antimicrobial resistance.

\section{Methods}
Recent developments in the biomedical applications of laser speckle imaging have catalyzed novel approaches to non-invasive diagnostic methods. The application of laser speckle in antibiogram analysis, particularly, has demonstrated significant potential for rapidly identifying bacterial species, which is critical for administering effective antimicrobial therapies \cite{pereira2022rapid}. Employing dynamic speckle patterns, these methods facilitate real-time detection and analysis of bacterial responses to antibiotics across diverse experimental settings, underscoring the need for further comprehensive testing to validate their efficacy.

In a broader diagnostic context, laser speckle techniques are diversely applied across three primary domains: phenotypic methods, molecular diagnostics, and mass spectrometry. Traditional phenotypic methods like disk diffusion and the quantitative MIC tests are valued for their reproducibility and remain benchmark techniques \cite{gajic2022antimicrobial, oberhettinger2020evaluation, dubourg2018rapid}. On the molecular front, innovative platforms such as the Biofire FilmArray and Genmark ePlex have considerably accelerated pathogen detection processes \cite{tansarli2020diagnostic, schmitz2018genmark}. Additionally, mass spectrometry approaches, particularly MALDI-TOF, have become indispensable for their swift and reliable species identification capabilities \cite{charretier2016mass, bizzini2010matrix}.

The versatility of laser speckle imaging has also been harnessed for 3D reconstruction applications that are becoming increasingly vital in surgical and dermatological settings. This technology enhances the dimensional perception of human tissue structures and aids in the early detection of melanocytic abnormalities \cite{khan2018single, gonzalez2018qualitative}. These applications not only demonstrate the broad utility of laser speckle imaging but also its capacity to refine surface characterization and improve diagnostic precision significantly.

This overview provides the necessary backdrop to introduce our study’s innovative use of laser speckle techniques in antibiogram performance enhancement. By focusing on addressing the conventional limitations, particularly in diagnostic delay, our research aims to advance the efficacy and responsiveness of microbial diagnostics.

Laser speckle is characterized by the interference pattern that forms when coherent light from a laser reflects off a surface rougher than the light's wavelength. This pattern comprises speckle grains, which are bright and dark spots resulting from constructive and destructive interference. The size of these grains depends on the light’s wavelength, the laser beam's diameter, and the distance from the surface to the observation plane. Fluctuations in the refractive index cause these speckle patterns to vary over time, which can be captured using a camera over a defined exposure period, providing dynamic insights into the speckle behavior \cite{dainty2013laser}.

\begin{figure}[ht]
\centering
\includegraphics[width=0.4\textwidth]{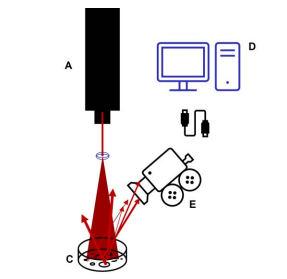}
\caption{Acquisition setup: A – Laser source, B – Lens; C – Sample; D – Computer; E – Video system (adapted from \cite{pereira2022rapid}).}
\label{fig:setup}
\end{figure}

The experimental setup includes essential components such as a camera, laser emitter, lens, and a Petri dish for holding biological samples. This setup facilitates both the acquisition and processing of speckle images, essential for analyzing surface interactions and biological responses. For image capture, a Basler acA1920-25gm camera is used, capable of recording at 25 FPS in Full HD resolution. In this case, the laser employed was a diode laser emitting at a wavelength of 650 nm with a power output of 5 mW. The acquisition setup, as shown in Figure \ref{fig:setup}, is powered via Power over Ethernet (PoE). This is facilitated by a NETGEAR GS110TP switch, which supports high data transfer rates crucial for maintaining image quality at high frame rates \cite{Basler2015}.

Image processing involves segmentation and statistical analysis. Initially, the watershed segmentation algorithm is employed to isolate regions of interest (ROI) within the speckle patterns. Subsequently, principal component analysis (PCA) is used to reduce the dimensionality of the data. This reduction is crucial for managing the complexity of the data and enhancing the efficiency of the subsequent clustering phase. Clustering algorithms then categorize the data into groups based on similarities in the speckle patterns, which helps in identifying patterns correlated with specific biological phenomena or changes. This methodology enables a detailed analysis of laser speckle patterns, aiding in the exploration of new diagnostic or investigative avenues in biomedical research. By capturing and analyzing these patterns, researchers can infer changes in material properties or biological conditions, underscoring the technique's applicative scope in scientific studies \cite{pereira2022rapid, khan2018single}.

\begin{figure}[ht!]
    \centering
    \includegraphics[width=0.8\textwidth]{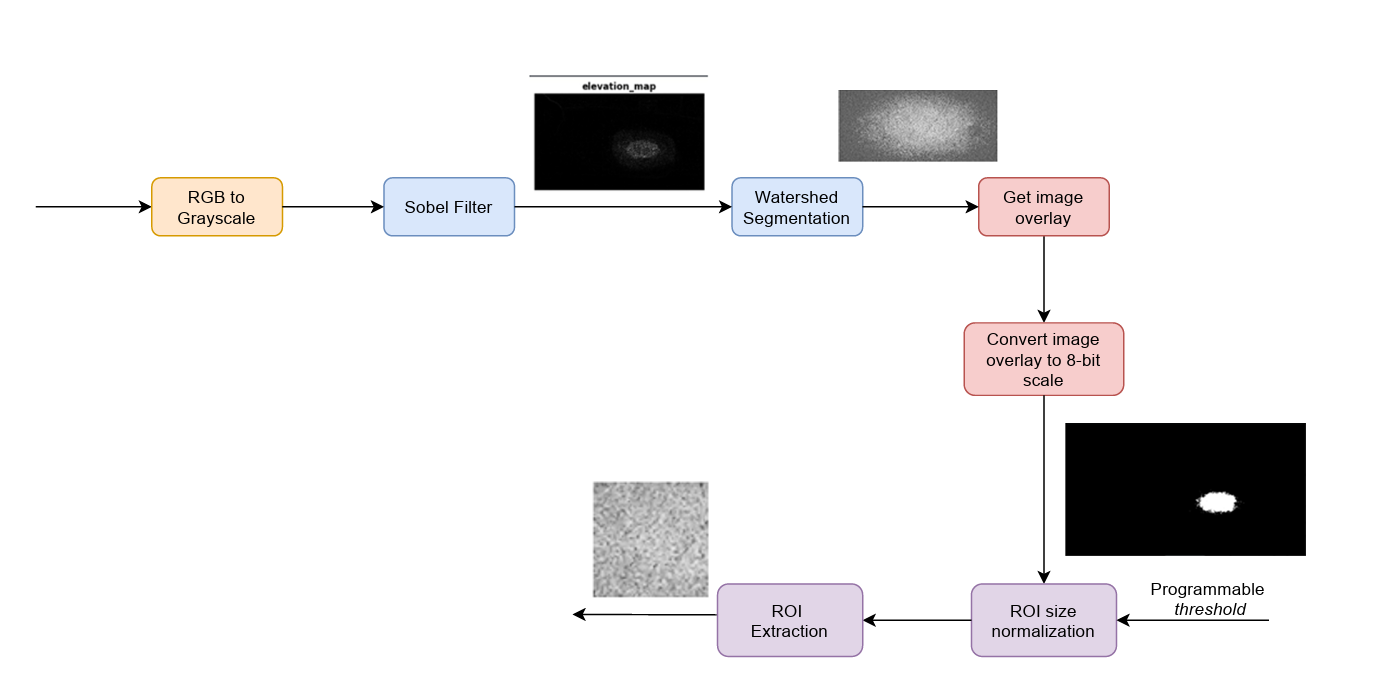}
    \caption{Procedure for the extraction of Regions of Interest (ROIs).}
    \label{fig:ROI_Extraction}
\end{figure}

Regarding the procedure for obtaining the ROIs from the acquired images, it is represented in Figure \ref{fig:ROI_Extraction}.

\begin{figure}[h!t]
    \centering
    \includegraphics[width=0.6\textwidth]{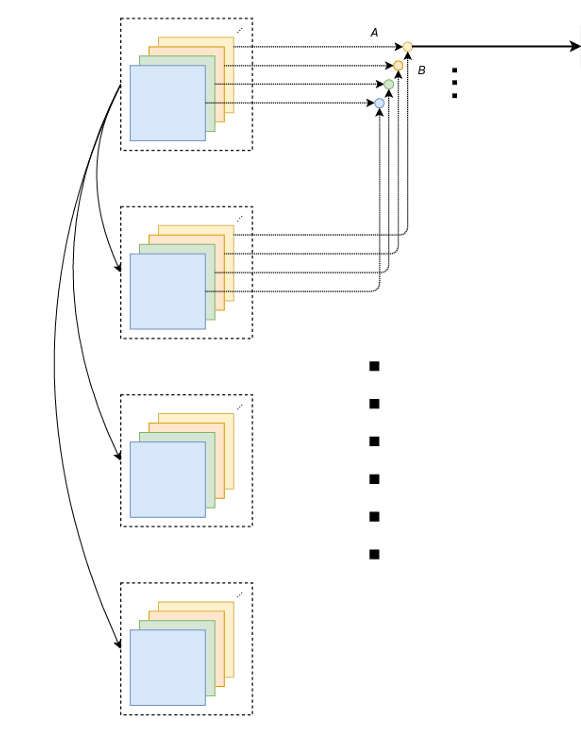}
    \caption{Procedure before the application of the features.}
    \label{fig:Features}
\end{figure}

The procedure before the application of the features is shown in the Figure \ref{fig:Features}. Each square represents a Region of Interest (ROI), and each set of squares corresponds to all the ROIs from the images within a single video. The contrast metric is calculated iteratively, comparing the ROI from the first image of the first video with the ROI from the first image of the second video, then the ROI from the second image of the first video with the ROI from the second image of the second video, and so on for each image and video pair.

%\onecolumn

\vspace{15cm}

%\twocolumn

\section{Results}

\begin{figure}[h!t]
\centering
\includegraphics[width=0.4\textwidth]{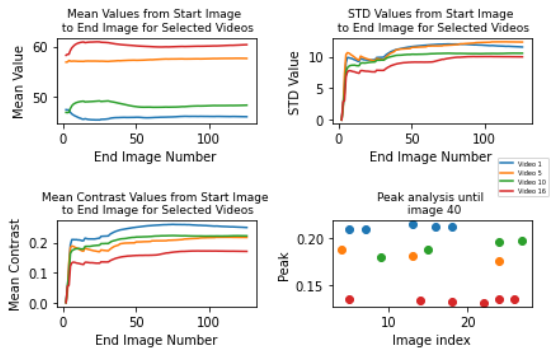}
\caption{Overview of the speckle contrast metrics over time, illustrating the decrease in contrast and stabilization, correlating with the phases of bacterial activity observed during the experiment.}
\label{fig:results}
\end{figure}

The experimental findings from the developed acquisition system and processing methodology are effectively summarized in Figure \ref{fig:results}. The experiment consisted of a total of 16 tests, with each test comprising a 5-second video recorded at 15-minute intervals over a total duration of 4 hours. The graph illustrates a notable decrease in speckle contrast over time, signifying an increase in bacterial activity, particularly pronounced within the first hour. This rapid change highlights the efficacy of speckle contrast as a metric for monitoring bacterial dynamics in real-time.

The speckle contrast metrics from the initial 10 frames exhibit substantial fluctuations, indicating significant bacterial activity early in the experiment. As the experiment progresses, these fluctuations stabilize, underscoring the dynamic nature of bacterial growth and response under controlled conditions.

\vspace{4cm}

\section{Discussion}
However, the PCA+Clustering methodology could extend total processing times, suggesting potential improvements through more traditional processing approaches, which were preliminarily tested in a controlled laboratory setting.

Looking ahead, it would be beneficial to make the Python-based software developed during this project open-source, and also enabling execution as executable files (.exe). This would widen the software’s accessibility and utility across the scientific and research communities. Implementing containerization technologies such as Docker could also enhance the project's compatibility with various operating systems, broadening its application. Furthermore, an extension of the experimental setup to include multiple lasers with different wavelengths and powers, and designing a system that accommodates liquid culture samples, could significantly expand the research's scope and applications.

\section{Conclusion}

The primary objective of this study was to develop an acquisition and processing system to monitor bacterial activity over time using laser speckle technology. The findings confirm that the contrast metric is inversely proportional to the elapsed time during the acquisition experiments, indicating that a decrease in contrast is directly associated with an increase in bacterial activity. Additionally, the imaging quality remained consistent throughout the experiments, underscoring the effectiveness of the experimental setup in maintaining image quality without significant variations in brightness.

The system proved to be versatile enough to allow adjustments in acquisition parameters, facilitating comprehensive testing with cultured samples both with and without antibiotic treatments. The flexibility in terms of recording times and the number of videos analyzed was successfully demonstrated.

\section*{Declarations}

\subsection*{Funding}
This study was funded by Funda\c{c}\~ao para a Ci\^encia e a Tecnologia (FCT).

\subsection*{Conflict of Interest}
The authors have no relevant financial or non-financial interests to disclose.

\subsection*{Data Availability Statement}
All data generated or analyzed during this study are available upon reasonable request from the corresponding author.

\subsection*{Authors' Contributions}
M.A. Gameiro, R.F. Pinto, and N.V. Lopes contributed equally to the design, execution, and analysis of this study. All authors reviewed and approved the final manuscript.

\subsection*{Acknowledgements}
I acknowledge the support from the Funda\c{c}\~ao para a Ci\^encia e a Tecnologia (FCT) for the research grant. Special thanks to Professors Rui Fonseca-Pinto and Nuno Vieira Lopes for their guidance and to my family and colleagues for their continuous support.

\bibliographystyle{apalike}
\bibliography{biblio}

\begin{thebibliography}{}

\bibitem[{Basler ace}, 2015]{Basler2015}
{Basler ace} (2015).
\newblock {\em User's Manual for GigE Cameras}.
\newblock Available: \url{https://www.baslerweb.com/en/sales-support/downloads/document-downloads/}.

\bibitem[Bereket et~al., 2012]{bereket2012update}
Bereket, W., Hemalatha, K., Getenet, B., Wondwossen, T., Solomon, A., Zeynudin, A., and Kannan, S. (2012).
\newblock Update on bacterial nosocomial infections.
\newblock {\em European Review for Medical \& Pharmacological Sciences}, 16(8).

\bibitem[Bizzini and Greub, 2010]{bizzini2010matrix}
Bizzini, A. and Greub, G. (2010).
\newblock Matrix-assisted laser desorption ionization time-of-flight mass spectrometry, a revolution in clinical microbial identification.
\newblock {\em Clinical Microbiology and infection}, 16(11):1614--1619.

\bibitem[Brusselaers et~al., 2011]{brusselaers2011rising}
Brusselaers, N., Vogelaers, D., and Blot, S. (2011).
\newblock The rising problem of antimicrobial resistance in the intensive care unit.
\newblock {\em Annals of intensive care}, 1:1--7.

\bibitem[Charretier and Schrenzel, 2016]{charretier2016mass}
Charretier, Y. and Schrenzel, J. (2016).
\newblock Mass spectrometry methods for predicting antibiotic resistance.
\newblock {\em PROTEOMICS--Clinical Applications}, 10(9-10):964--981.

\bibitem[Dainty, 2013]{dainty2013laser}
Dainty, J.~C. (2013).
\newblock {\em Laser speckle and related phenomena}, volume~9.
\newblock Springer science \& business Media.

\bibitem[Dubourg et~al., 2018]{dubourg2018rapid}
Dubourg, G., Lamy, B., and Ruimy, R. (2018).
\newblock Rapid phenotypic methods to improve the diagnosis of bacterial bloodstream infections: meeting the challenge to reduce the time to result.
\newblock {\em Clinical Microbiology and Infection}, 24(9):935--943.

\bibitem[Gajic et~al., 2022]{gajic2022antimicrobial}
Gajic, I., Kabic, J., Kekic, D., Jovicevic, M., Milenkovic, M., Mitic~Culafic, D., Trudic, A., Ranin, L., and Opavski, N. (2022).
\newblock Antimicrobial susceptibility testing: a comprehensive review of currently used methods.
\newblock {\em Antibiotics}, 11(4):427.

\bibitem[Gamer et~al., 1988]{gamer1988cdc}
Gamer, J., Jarvis, W., Emori, T., Horan, T., and Hughes, J. (1988).
\newblock Cdc definitions for nosocomial infections.
\newblock {\em Am J Infect Control}, 16(3):128--40.

\bibitem[Gonzalez et~al., 2018]{gonzalez2018qualitative}
Gonzalez, M.~A., Guzm{\'a}n, M.~N., Fonseca-Pinto, R., Trivi, M., Rabal, H., and Passoni, L.~I. (2018).
\newblock Qualitative characterization of skin tissue with dynamic laser speckle.
\newblock {\em Revista Argentina de Bioingenier{\'\i}a}, pages 19--23.

\bibitem[Khan et~al., 2018]{khan2018single}
Khan, D., Shirazi, M.~A., and Kim, M.~Y. (2018).
\newblock Single shot laser speckle based 3d acquisition system for medical applications.
\newblock {\em Optics and Lasers in Engineering}, 105:43--53.

\bibitem[Oberhettinger et~al., 2020]{oberhettinger2020evaluation}
Oberhettinger, P., Zieger, J., Autenrieth, I., Marschal, M., and Peter, S. (2020).
\newblock Evaluation of two rapid molecular test systems to establish an algorithm for fast identification of bacterial pathogens from positive blood cultures.
\newblock {\em European Journal of Clinical Microbiology \& Infectious Diseases}, 39:1147--1157.

\bibitem[Pereira et~al., 2022]{pereira2022rapid}
Pereira, S.~G., Verdugo, J., and Fonseca-Pinto, R. (2022).
\newblock Rapid antimicrobial susceptibility testing using laser speckle technology.
\newblock In {\em 2022 45th Jubilee International Convention on Information, Communication and Electronic Technology (MIPRO)}, pages 389--392. IEEE.

\bibitem[Schmitz and Tang, 2018]{schmitz2018genmark}
Schmitz, J.~E. and Tang, Y.-W. (2018).
\newblock The genmark eplex{\textregistered}: another weapon in the syndromic arsenal for infection diagnosis.
\newblock {\em Future microbiology}, 13(16):1697--1708.

\bibitem[Tansarli and Chapin, 2020]{tansarli2020diagnostic}
Tansarli, S. and Chapin, C. (2020).
\newblock Diagnostic test accuracy of the biofire{\textregistered} filmarray{\textregistered} meningitis/encephalitis panel: a systematic review and meta-analysis.
\newblock {\em Clinical Microbiology and Infection}, 26(3):281--290.

\bibitem[Walsh et~al., 2023]{walsh2023antimicrobial}
Walsh, T.~R., Gales, A.~C., Laxminarayan, R., and Dodd, P.~C. (2023).
\newblock Antimicrobial resistance: addressing a global threat to humanity.

\end{thebibliography}

\end{document}